\documentclass[usenatbib,useAMS,letters]{mnras}%
\usepackage{tablefootnote}
\usepackage{pdflscape}
\usepackage{multirow}
\usepackage{xcolor}
\usepackage{hyperref}
\usepackage{graphicx}	
\usepackage{amsmath}	
\usepackage{amssymb}	
\usepackage{multicol}        
\usepackage{bm}		
\usepackage[normalem]{ulem} 
\usepackage{caption}      
\usepackage{url}
\usepackage{longtable}

\title[Catalogue of blazars]{The RATAN-600 multi-frequency catalogue of blazars -- BLcat}

\author[Yu. Sotnikova et al.]{
Yu. Sotnikova,$^{1}$\thanks{E-mail: lacertae999@gmail.com} 
T.~Mufakharov,$^{1,2}$
R.~Udovitskiy,$^{1}$
M.~Mingaliev,$^{1,2,3}$
T. Semenova,$^{1}$
A. Erkenov,$^{1}$
\newauthor
N.~Bursov,$^{1}$
A. Mikhailov,$^{1}$
Yu. Cherepkova$^{1}$
\\
$^{1}$Special Astrophysical Observatory of RAS, Nizhny Arkhyz, 369167 Russia\\
$^{2}$Kazan Federal University, Kazan, 420008 Russia\\
$^{3}$Institute of Applied Astronomy of RAS, St. Petersburg, 191187 Russia
}

\date{Received 17 August 2022. Accepted 9 September 2022. \\ {\bf Published in Astrophysical Bulletin Vol.~77, No.4, pp.397--410 (2022)} \\\textbf{DOI:} 10.1134/S1990341322040149}
\pubyear{2022}
\begin{document}
\label{firstpage}
\pagerange{\pageref{firstpage}--\pageref{lastpage}}
\maketitle 
  
\begin{abstract} 
In this paper we present the RATAN-600 multi-frequency catalogue of blazars, an updated version of the BLcat: the RATAN-600 multi-frequency catalogue of BL Lacertae objects. The main novelty in the catalogue is an extension of the sample with flat-spectrum radio quasars (FSRQs), thus currently it contains more than 1700~blazars of different types. The main feature of the BLcat is a compilation of radio continuum data for blazars based on the RATAN-600 quasi-simultaneous measurements at frequencies of 1.2, 2.3, 4.7, 7.7/8.2, 11.2, and 21.7/22.3~GHz. We additionally supplement the catalogue with the radio data from external sources to provide an opportunity to more complete study of radio spectra and radio light curves. For the convenience of users, we developed tools to calculate the spectral index, variability index, and radio luminosity. We briefly describe basic radio properties of blazar subsamples of the catalogue: spectral classification, spectral indices, flux density variability, and radio luminosity.
\end{abstract}



\section{Introduction}
\label{introduction}
\begin{figure}
\centerline{\includegraphics[width=\columnwidth]{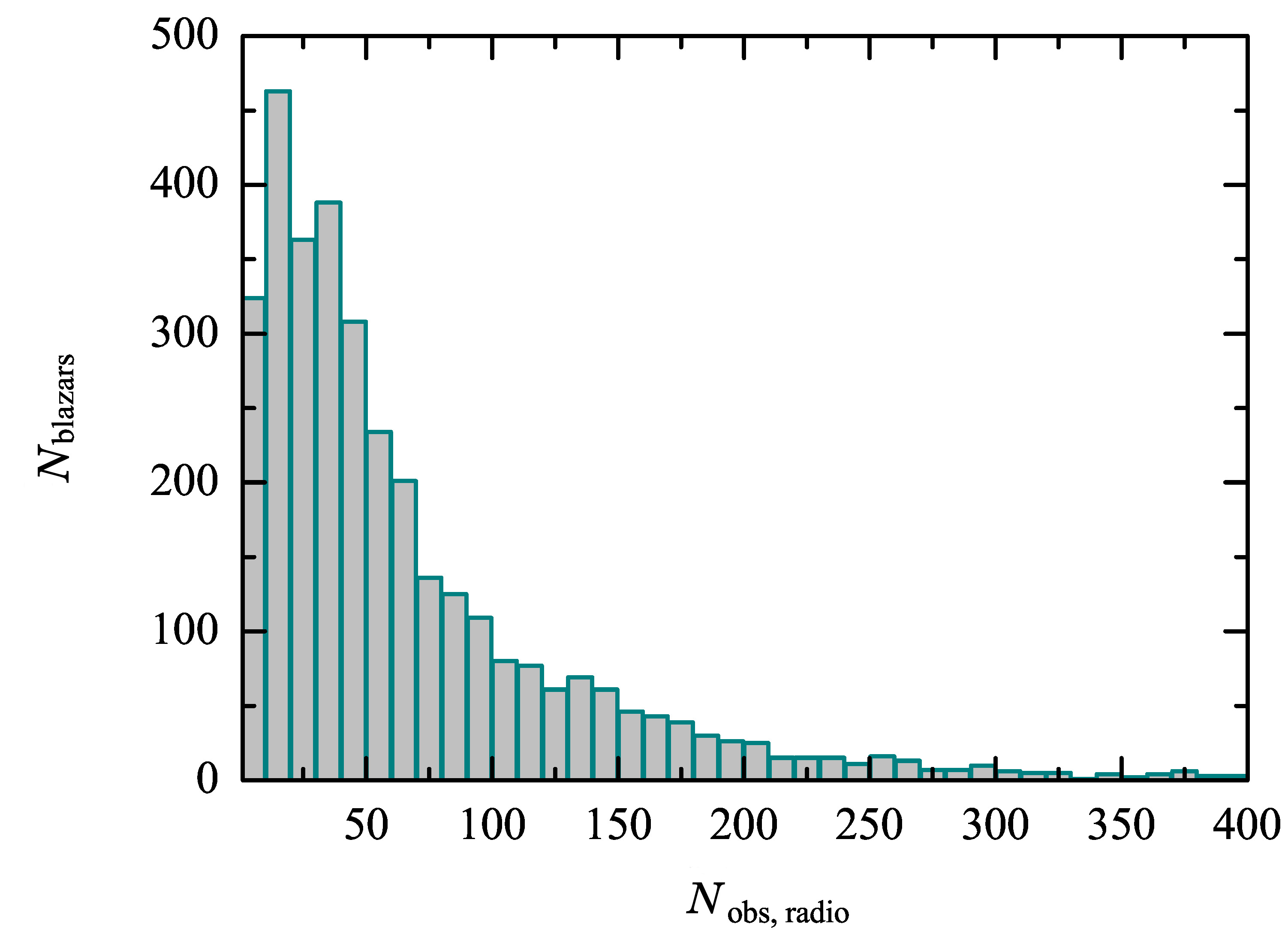}}
\caption{The number of literature radio measurements for the Roma-BZCAT blazars (3561~objects). About 200 blazars with more than 400 measurements were excluded for a clear view.}
\label{fig1}
\end{figure}

Blazars are radio-loud active galactic nuclei (AGNs) with their jets pointed in the direction to the observer \citep{1995PASP..107..803U,2017A&ARv..25....2P}. After their discovery, they became interesting as the objects of extreme physical conditions with rapid and violent variability at different timescales. Later, in addition to that, the interest to their investigation was triggered by the fact that they are the most common sources of $\gamma$-ray emission in the sky \citep{2020ApJ...892..105A}. Recently, they were discovered as the possible sources of PeV-energy neutrino \citep{2018Sci...361..147I}. For a comprehensive study of blazars and the connection between the radio emission and neutrinos in them, it is important to conduct multi-band high-cadence long-term observations \citep{2020AdSpR..65..745K,2020ApJ...894..101P,2021A&A...650A..83H}. The Whole Earth Blazar Telescope (WEBT, \citealt{2002MmSAI..73.1191V}) is a collaboration which organises such campaigns on blazar monitoring, from radio to $\gamma$-rays. In the radio bands, the greatest contribution was made by several observatories: the long-term regular monitoring of brightest AGNs at 37~GHz by the Mets{\"a}hovi Radio Observatory \citep{1987A&AS...70..409S,2009AJ....137.5022N}, the 15~GHz blazar monitoring program by the Owens Valley Radio Observatory \citep{2011ApJS..194...29R}, the three-frequency (4.8--14.5~GHz) monitoring by the University of Michigan Radio Astronomy Observatory \citep{1985ApJS...59..513A}, the 1.4~GHz NRAO VLA Sky Survey (NVSS, \citealt{1998AJ....115.1693C}), the 8.4~GHz observations with the Very Large Array (VLA, \citealt{2007ApJS..171...61H}), and the 2--4~GHz Very Large Array Sky Survey (VLASS, \citealt{2020PASP..132c5001L}). However, long-term light curves are usually sparsely sampled, and most blazars have been observed mainly 
within multi-frequency flare activities monitoring campaigns \citep{2019NewAR..8701541H}. For example, in the Roma-BZCAT Multi-frequency Catalogue of Blazars \citep{2015Ap&SS.357...75M} more than 20\% of them have less than 20 radio measurements (Fig.~\ref{fig1}).

Among the mentioned radio telescopes and surveys, the RATAN-600 radio telescope stands out for its capability of obtaining six-frequency measurements (1--22~GHz) simultaneously. AGNs, including blazars, have been regularly monitored with RATAN-600 since 2005, and this data formed the basis for the first version of the BLcat\footnote{\url{https://www.sao.ru/blcat/}} \citep{2014A&A...572A..59M}, the catalogue of ``RATAN-600 multi-frequency data for the BL Lacertae objects''. It contained more than 300 BL Lac objects and candidates. In this paper, we present the electronic ``RATAN-600 multi-frequency catalogue of blazars'', which is an update of the BLcat. In this new BLcat edition~1.3 (September 2021), we present several major changes to the previous version: firstly, the catalogue was enlarged with new types of blazars, namely blazars of an uncertain type (Blaz.un.t) and flat-spectrum radio quasars (FSRQs), and now it contains more than 1700 blazars; secondly, the BL Lac objects from the first version of the catalogue were supplemented by new RATAN-600 flux density measurements; and thirdly, the radio data from external databases and the literature are given along with the RATAN-600 data. 

The paper is organised as follows. The properties of the sample are given in Section~2; the description of the \mbox{RATAN-600} observations and characteristics of the antenna and receivers are given in Section~3.1; in Section~3.2 the external data added to the catalogue are described; in Section~4 we present the general organisation, the structure of the catalogue, and available tools. Section~5 describes radio properties of blazar populations of the catalogue. The summary of the paper is given in Section~6.

\section{The sample}
\label{sample}
The list of BLcat blazars is based on the Roma-BZCAT catalogue by \citet{2009A&A...495..691M}, which is the largest collection of blazars selected from different multi-band surveys. Its 5th edition \citep{2015Ap&SS.357...75M} contains 3561 objects, all of them provided with the radio band data from the NVSS \citep{1998AJ....115.1693C}, the FIRST survey \citep{1997ApJ...475..479W}, or the SUMMS \citep{2003MNRAS.342.1117M}. The BZCAT blazar classification is based in general on the spectral energy distribution (SED) showing strong non-thermal emission over the entire electromagnetic spectrum and the presence of evidence for relativistic beaming. When available, the data for $\gamma$-rays, soft and hard X-rays, microwaves, and the $R$-band magnitudes are also given. 

The BLcat catalogue originally included 306 BL Lacertae objects from the large Mets{\"a}hovi BL Lac sample \citep{2006A&A...445..441N} and contained the RATAN-600 measurements in 2005--2014 \citep{2014A&A...572A..59M,2017AN....338..700M}. Later we included to the catalogue the FSRQ blazars from Roma-BZCAT within the declination range from $-34\degr$ to $+49\degr$ with an NVSS flux density limit of $S_{1.4} \geq 100$ mJy. In comparison with the first version of the catalogue \citep{2014A&A...572A..59M} current edition contains almost six times more objects, the population is mainly extended because of FSRQ type of blazars (Table~\ref{table1}). 

\begin{figure}
\centerline{\includegraphics[width=\columnwidth]{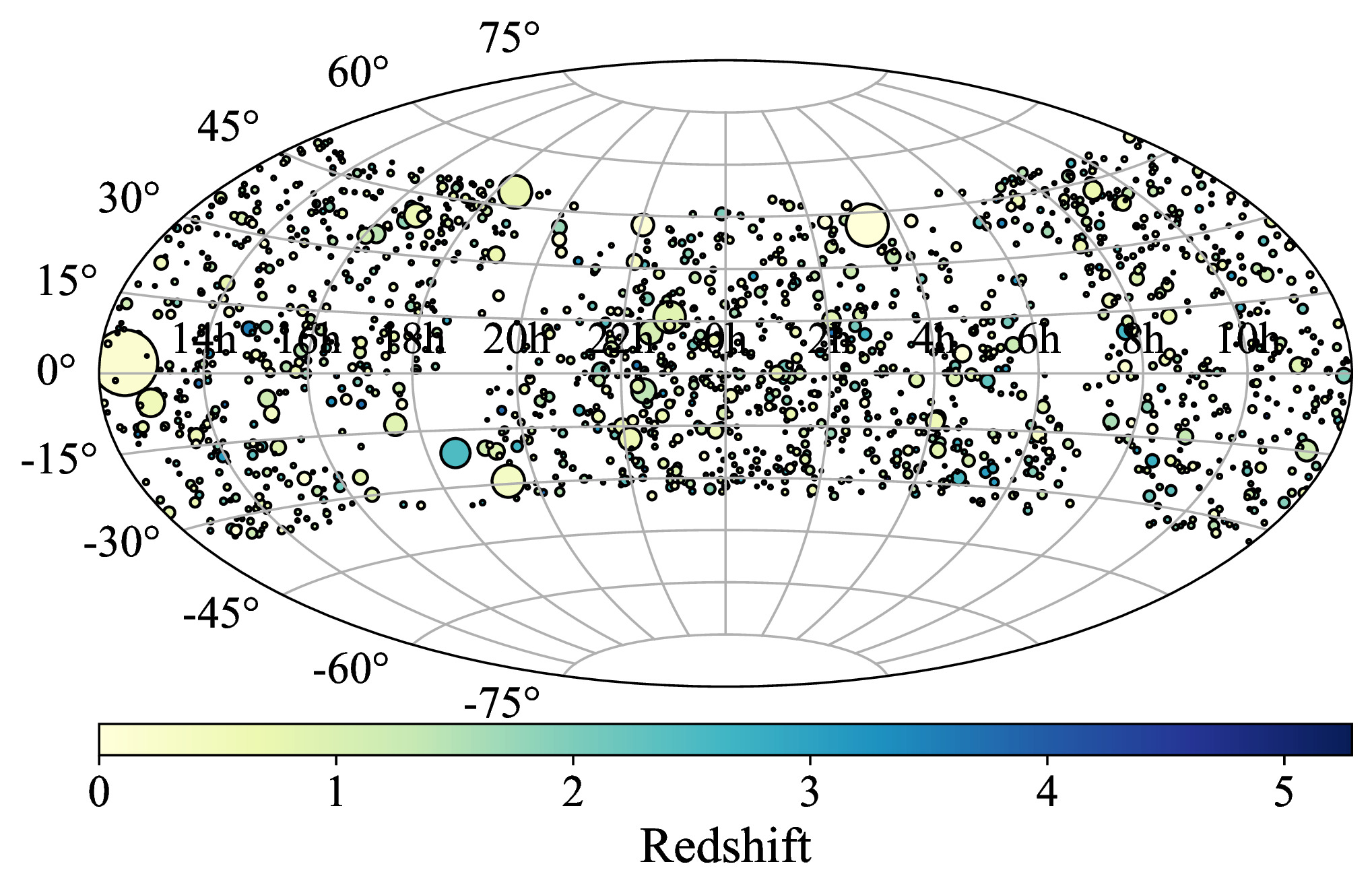}}
\caption{The Hammer--Aitoff projection (in equatorial coordinates) of the sky distribution of the objects from the BLcat catalogue. The size of the circles corresponds to the flux density level at 1.4~GHz, and the colour indicates the redshift value.}
\label{fig2}
\end{figure}

\begin{figure}
\centerline{\includegraphics[width=\columnwidth]{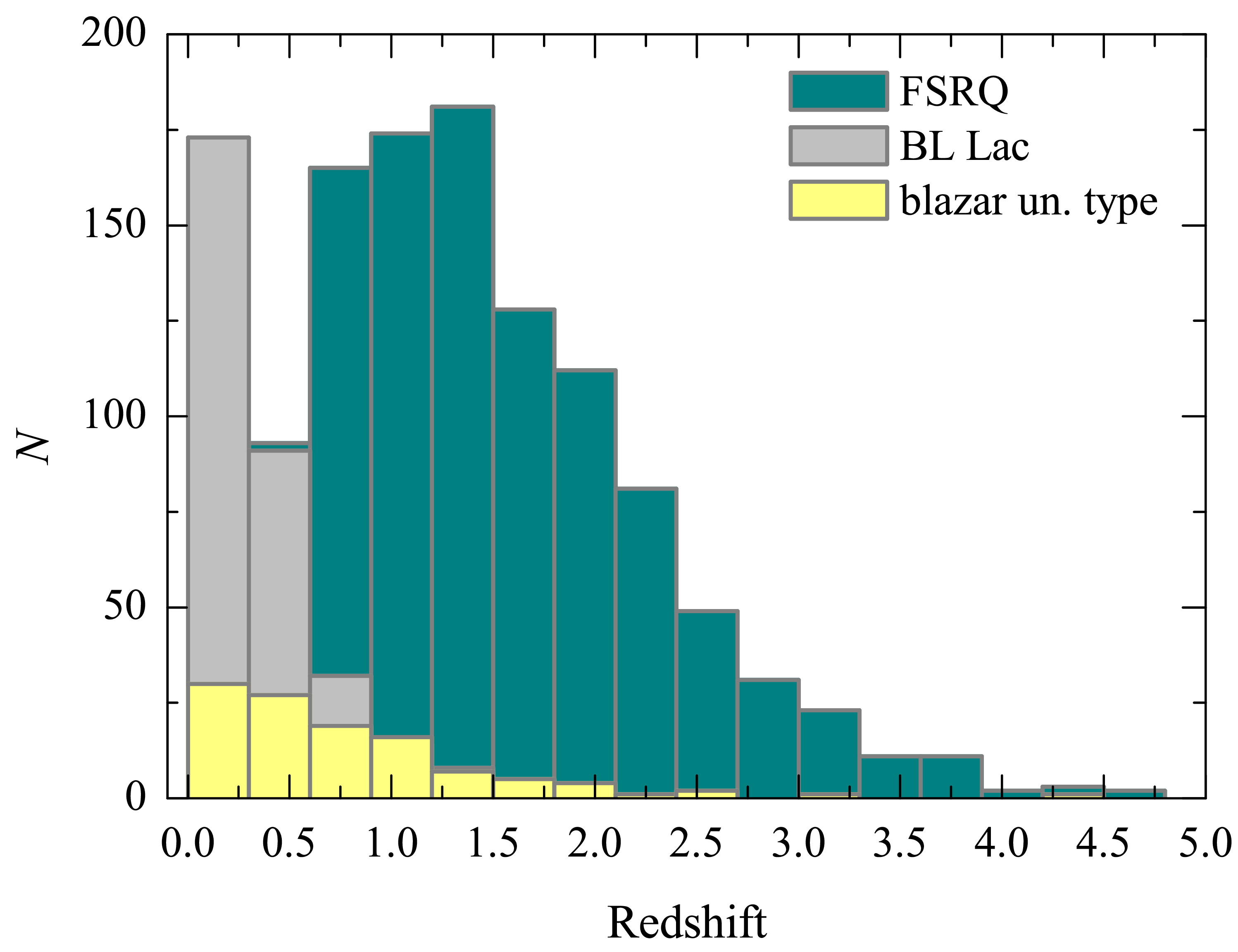}}
\caption{Redshift distribution. FSRQs are marked by green colour, BL Lac's are grey, and blazars of uncertain type are marked by yellow colour.}
\label{fig3}
\end{figure}

The BLcat aims to fully represent the RATAN-600 long-term measurements of blazars; an external radio data are being systematically included in the catalogue. At present, the catalogue contains more than 1700 blazars, which is almost half of all the Roma-BZCAT sources. We adopted the blazar classification, redshifts, and $R$-band magnitudes from Roma-BZCAT. The BL Lacertae are subclassified as BL~Lacertae--type objects, BL Lac candidates, and BL~Lac--galaxy dominated sources according to Roma-BZCAT. Fig.~\ref{fig2} shows the distribution of the objects from the catalogue in the sky, with the redshifts designated by the colour, and the source's flux densities shown by the marker size.

The redshift distribution for different blazar populations is shown in Fig.~\ref{fig3}. The median value of $z$ is 1.34 for FSRQs, 0.58 for blazars uncertain type, and 0.27 for BL Lacs. FSRQs exhibit larger scatter in a range of redshift, from 0.09 to 5.29.

\begin{table}
\caption{\label{table1} Blazars subclasses in the first and in the current editions of the BLcat. N - is the number of sources.}
\centering
\begin{tabular}{l|rl|rl}
\hline
\multicolumn{1}{c|}{\multirow{2}{*}{Class}} & 
\multicolumn{4}{c}{$N$ (\%)} \\
\cline{2-5}  
                                            & \multicolumn{2}{c|}{2014}              & \multicolumn{2}{c}{2022}     \\ \hline
FSRQ                                        & 7             & (2.3)                  & 1097& (62.2)                  \\ 
BL Lac                                      & 220           & (71.9)                 & 416 & (23.6)                  \\ 
BL Lac candidates                           & 43            & (14.1)                 & 32  &  (1.8)                  \\
BL Lac galaxy-dominated                     & \multicolumn{2}{c|}{\multirow{1}{*}{--}}& 92  &  (5.2)                  \\ 
Blazars of uncertain type                   & 36            & (11.8)                 & 128 &  (7.2)                  \\\hline
All                                         & 306           &                        &1765 &                         \\
\hline
\end{tabular}
\end{table}

\section{CATALOGUE'S CONTENTS}

\subsection{RATAN-600 Data}
\label{observations}
The observations of blazars with RATAN-600 have been carried out since 2005, and part of the measurements were published in the electronic catalogues   \citet{2008yCat..80840387M,2012yCat..35440025M,2015yCatp033007003M,2015yCat..74502658M,2015yCatp033006903M,2017yCat.113380700M,2019yCatp033007403S}. RATAN-600 radio continuum spectra are measured simultaneously at several frequencies on a time scale of 3--5 minutes \citep{1979S&T....57..324K,1993IAPM...35....7P}. We used two radiometric complexes working at three (4.7, 11.2, 21.7/22.3~GHz) and six (1.2, 2.3, 4.7, 7.7/8.2, 11.2, 21.7/22.3~GHz) frequencies \citep{2011AstBu..66..109T,2018AstBu..73..494T}. The parameters of the radiometers and RATAN antenna system are given in Table~\ref{table2}. The observations were processed using the automated data reduction system \citep{2016AstBu..71..496U}, which is based on the Flexible Astronomical Data Processing System (FADPS) standard data reduction software developed by \citet{1997ASPC..125...46V}.

The following flux density secondary calibrators were used: 3C48, 3C138, 3C147, 3C161, 3C286, 3C295, 3C309.1, and NGC~7027. The flux density scale is based on the measurements by \citet{1977A&A....61...99B} and \citet{2013ApJS..204...19P,2017ApJS..230....7P}, which are in good agreement with each other and differ only within the measurement errors. We also used the traditional RATAN-600 flux density calibrators: J0240$-$23, J1154$-$35, J1347$+$12, and J0521$+$16 \citep{2019AstBu..74..497S}. The measurements of the calibrators were corrected for angular size and linear polarisation according to the data from \citet{1994A&A...284..331O} and \citet{1980A&AS...39..379T}.

A standard error of the flux density includes an uncertainty of the RATAN-600 calibration curve (about 2--10\%) and the antenna temperature measurement error. The systematic uncertainty of the absolute flux density scale, which is 3--10\% at different frequencies \citep{1977A&A....61...99B}, is excluded in the total error. The median values of the standard error are 3--10\% at 11.2, 8.2, and 4.7~GHz, and 7--20\% at 2.3, 1.2, and 22.3~GHz.

At present, the number of RATAN observations is non-uniform over the whole BLcat list, it varies from 1 up to more than 100. The average number of observations is 17. Most blazars, about 65\%, have been observed on average 5 times. About 10\% of blazars have 15--45 RATAN-600 observational epochs; 22\% -- 45--70 epochs, and 4\% -- 70--115 epochs.

\begin{table}
\caption{The RATAN-600 continuum radiometer parameters: the central frequency $f_0$, the bandwidth $\Delta f_0$, the detection limit for point sources per transit $\Delta F$. ${FWHM}_{\rm {RA} \times \rm {Dec}}$ is the angular resolution along RA and Dec. calculated for {${\rm Dec} = 0\degr$}} 
\label{table2}
\centering
\begin{tabular}{c|c|c|@{~~~}c@{$~~\times$}@{\!\!\!\!}c}
\hline
$f_{0}$, & $\Delta f_{0}$,  & $\Delta F$, & \multicolumn{2}{c}{$FWHM_{\rm {RA} \times \rm{Dec}}$}\\
GHz     &   GHz           &  mJy/beam  &    \\
\hline
 $21.7/22.3$ & $2.5$  &  $50$ & $0\farcm17$ & $1\farcm8$  \\ 
 $11.2$ & $1.4$  &  $15$ & $0\farcm34$ &$3\farcm9$ \\ 
 $7.7/8.2$  & $1.0$  &  $10$ & $0\farcm47$ & $5\farcm2$   \\ 
 $4.7$  & $0.6$  &  $8$  & $0\farcm85$ &$9\farcm5$   \\ 
 $2.25$  & $0.08$  &  $40$ & $1\farcm64$ & $19\farcm2$  \\ 
 $1.25$  & $0.08$ &  $200$ & $3\farcm0$ & $32\farcm4$ \\ 
\hline
\end{tabular}
\end{table} 

\subsection{External Data}
\label{data}
\begin{figure}
\centerline{\includegraphics[width=\columnwidth]{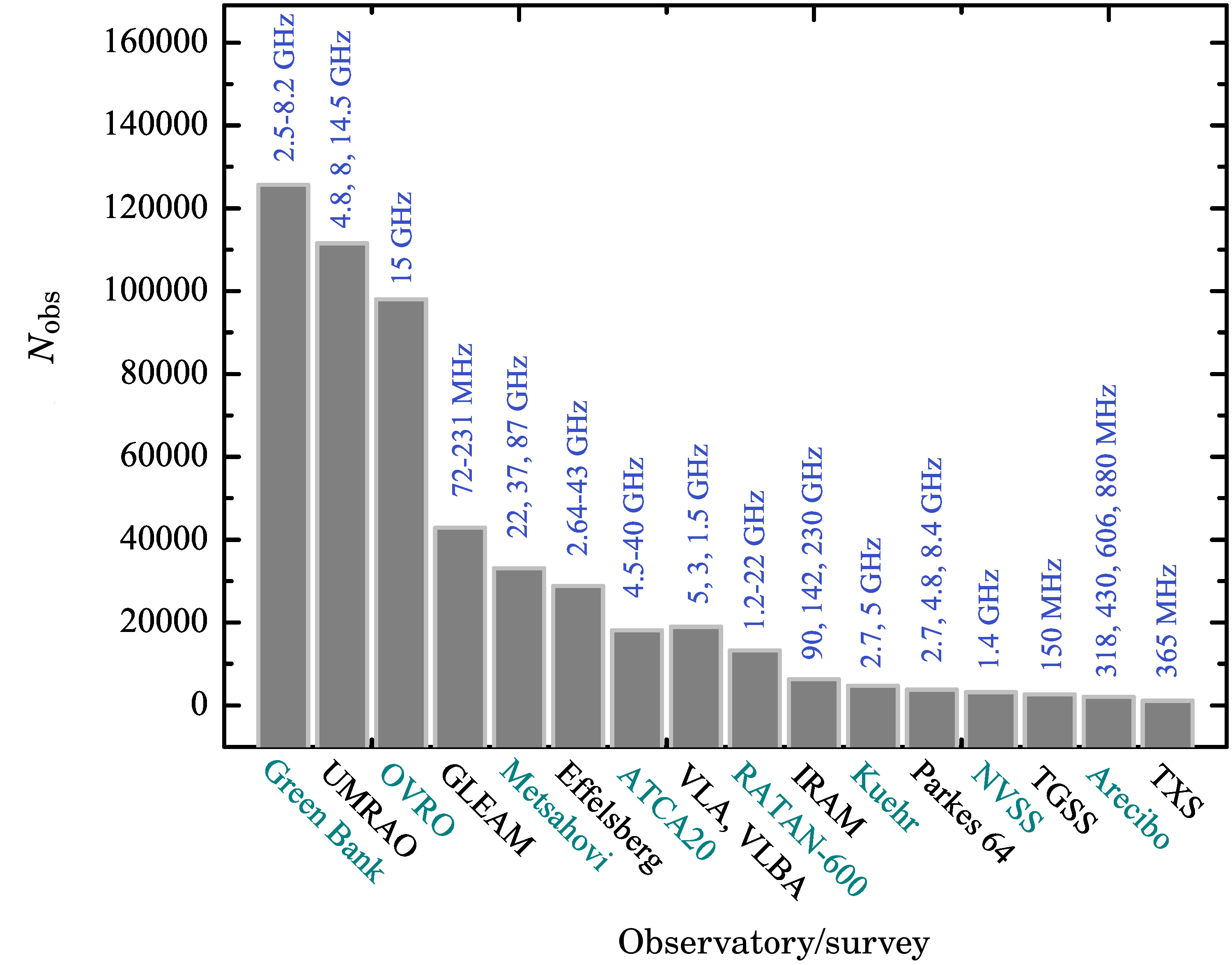}}
\caption{The surveys and observatories with the most contribution to the blazar radio continuum data in BLcat.}
\label{fig4}
\end{figure}

The new BLcat edition collects external radio continuum data from CATS\footnote{\url{https://www.sao.ru/cats/}}, the Astrophysical CATalogs support System database \citep{1997BaltA...6..275V,2005BSAO...58..118V,2009DatSJ...8...34V}, available on the Special Astrophysical Observatory website, from NED\footnote{\url{http://ned.ipac.caltech.edu}}, the NASA/IPAC Extragalactic Database, and from the VizieR Information System\footnote{\url{https://vizier.u-strasbg.fr/viz-bin/VizieR}} \citep{2000A&AS..143...23O}. The data used for the construction of broadband radio continuum spectra and cover the frequency range from tens of MHz up to hundreds of GHz and time period of 30--40~years. The total list with references is given in Table~\ref{table_C}. Fig.~\ref{fig4} presents the distribution of main of surveys or monitoring programs contributing to the blazar radio data. 

\begin{table*}
\caption{\label{table_C} External catalogs and programs with radio data included in the current edition}
\centering
\begin{tabular}{c|c|c|c}
\hline
Catalog/survey & Observatory/telescope & Frequency range, GHz & Refs\\
\hline
87GB, GB6&GBT&2.5, 4.85, 8.2&[1, 2, 3, 4] \\ 
&UMRAO&4.8, 8, 14.5&[5] \\ 
F-GAMMA&OVRO&10.8, 15&[6, 7] \\ 
GLEAM&MWA&0.072--0.231&[8] \\ 
F-GAMMA&Effelsberg&2.64--43&[9] \\ 
AT20G, PACO&ATCA&20, 4.5--40&[10, 11] \\ 
&Mets{\"a}hovi&22, 37, 87&[12, 13, 14] \\ 
VLASS&VLA&3&[15] \\ 
&VLBA&1.4, 5&[16] \\ 
&RATAN-600&1--22&[17, 18, 19, 20] \\ 
&IRAM&90, 142, 230&[21] \\ 
&GBT, Parkes&2.7, 5&[22] \\ 
NVSS&VLA&1.4&[23] \\ 
TGSS&GMRT&0.15&[24] \\ 
PMN&Parkes&4.85, 8.4&[25, 26] \\ 
&Arecibo&0.32, 0.43&[27] \\ 
&GBT& 0.61, 0.88, 1.4&[27] \\ 
TXS&UTRAO&0.365&[28] \\ 
\hline
\multicolumn{4}{p{12.5cm}}{Refs: [1, 2, 3, 4]~--- {\citet{2001ApJS..136..265L,1986A&AS...65..267A}},
\citet{1991ApJS...75.1011G,1991ApJS...75....1B}; [5]~--- \citet{1985ApJS...59..513A};
[6, 7]~--- \citet{2011ApJS..194...29R,1983PASP...95..842S}; [8]~--- \citet{2017MNRAS.464.1146H}; 
[9]~--- {\citet{2019A&A...626A..60A}}; [10, 11]~--- \citet{2011MNRAS.415.1597M,2008MNRAS.384..775M}; 
[12, 13, 14]~--- {\citet{1998A&AS..132..305T,1992A&AS...94..121T,2004A&A...427..769T}}; 
[15, 16]~--- \citet{2020RNAAS...4..175G,1996ApJS..107...37T}; 
[17, 18, 19, 20]~--- {\citet{2017AN....338..700M,2012A&A...544A..25M,
1999A&AS..139..545K, 2002BSAO...54....5K}}; 
[21]~--- {\citet{1997A&AS..122..271R}}; 
[22]~--- {\citet{1981A&AS...45..367K}}; 
[23]~--- \citet{1998AJ....115.1693C};
[24]~--- {\citet{2017A&A...598A..78I}}; 
[25, 26]~--- \citet{1990PASA....8..261W,1996APJS..103..145W}; 
[27]~--- \citet{1994ApJS...93..441M}; 
[28]~--- \citet{1996AJ....111.1945D}.}\\
\end{tabular}
\end{table*}

External data collected by different instruments are often inhomogenous. For example, interferometric observations have higher angular resolution than our RATAN-600 telescope, thus leading to underestimations of source's flux densities compared to RATAN-600 measurements. On the other hand, measurements from single-dish telescopes with beams larger than those of RATAN-600 may be impacted by confusion with nearby sources, leading to overestimates of source's flux densities. Noting this we listed and classified all the catalogues from CATS, separating interferometric from single-dish data. We marked interferometric and single-dish measurements with different colours and they are optionally allowed to be drawn in cumulative radio spectrum.

\begin{table*}
\caption{The first ten rows given as an example of the structure and information stored in the main table of the BLcat catalogue edition~1.3 (September 2021)}
\label{table3}
\centering
\begin{tabular}{c|c|c|c|c|c|c|c|c|c}
\hline
\multirow{2}{*}{$N$}       & \multirow{2}{*}{Epochs} & Source & RA       & Dec       & \multirow{2}{*}{$z$} & \multirow{2}{*}{R$_{\rm mag}$} & \multirow{2}{*}{$S_{4.7}$, Jy}  &  \multirow{2}{*}{$P_{4.7}$, W Hz$^{-1}$}  & Blazar\\
                         &     & name   & (J2000.0)& (J2000.0) &         & &  &  &  type    \\
\hline
(1)  &  (2)  &  (3) &  (4)&  (5) &    (6)   &  (7)&  (8)&  (9) &   (10)    \\
\hline
1 & 2  & 5BZQJ0001$-$1551 &  00:01:05 & $-$15:51:06 & 2.044 & 18.1 & 0.20 & $3.23\times10^{27}$ & FSRQ  \\ 
2 & 6  & 5BZQJ0001$+$1914 &  00:01:08 & $+$19:14:34 & 3.1   & 21.6 & 0.13 & $3.04\times10^{27}$ & FSRQ  \\ 
3 & 3  & 5BZBJ0001$-$0746 &  00:01:18 & $-$07:46:26 & --    & 17.9 & 0.18 & -- & BL\,Lac \\ 
4 & 6  & 5BZBJ0001$-$0011 &  00:01:21 & $-$00:11:39 & 0.462 & 19.6 & 0.05 & $3.26\times10^{25}$ & BL\,Lac \\ 
5 & 46 & 5BZBJ0002$-$0024 &  00:02:57 & $-$00:24:47 & 0.523 & 19.7 & 0.09 & $8.55\times10^{25}$ & BL\,Lac \\ 
6 & 1  & 5BZBJ0004$-$1148 &  00:04:04 & $-$11:48:57 & --    & 18.8 & 0.78 & -- & BL\,Lac \\ 
7 & 2  & 5BZQJ0004$+$4615 &  00:04:16 & $+$46:15:18 & 1.81  & 20.4 & 0.24 & $2.35\times10^{27}$ & FSRQ \\ 
8 & 2  & 5BZQJ0005$-$1648 &  00:05:17 & $-$16:48:04 & 0.78  & 18.4 & 0.15 & $3.01\times10^{26}$ & FSRQ  \\ 
9 & 1  & 5BZQJ0005$+$0524 &  00:05:20 & $+$05:24:10 & 1.9   & 16.2 & 0.14 & $1.39\times10^{27}$ & FSRQ  \\ 
10 & 66  & 5BZQJ0005$+$3820 &  00:05:57 & $+$38:20:15 & 0.229   & 17.6 & 0.50 & $6.35\times10^{25}$ & FSRQ  \\ 
\hline
\multicolumn{10}{p{16cm}}{Column designations: (1)---the entry number, (2)---the number of RATAN-600 observing epochs, (3)---the source name, (4)---right ascension (J\,2000, hh:mm:ss), (5)---declination (J\,2000,  dd:mm:ss), (6)---redshift, (7)---the $R$-band magnitude from USNO-B1, (8)---the average flux density at 4.7~GHz measured based on the RATAN-600 observations, (9)---the average radio luminosity at 4.7~GHz measured based on the RATAN-600 observations, (10)---the blazar type based on the BZCAT classification: BL~Lac, BL~Lac candidate (BL~Lac cand.), BL~Lac-galaxy dominated (BL~Lac.g.dom.), FSRQ, or a blazar of an uncertain type (Blaz.un.t.).}\\
\end{tabular}
\end{table*}

\section{General organisation of the catalogue}
\label{organization}

\begin{figure*}
\centerline{\includegraphics[width=\textwidth]{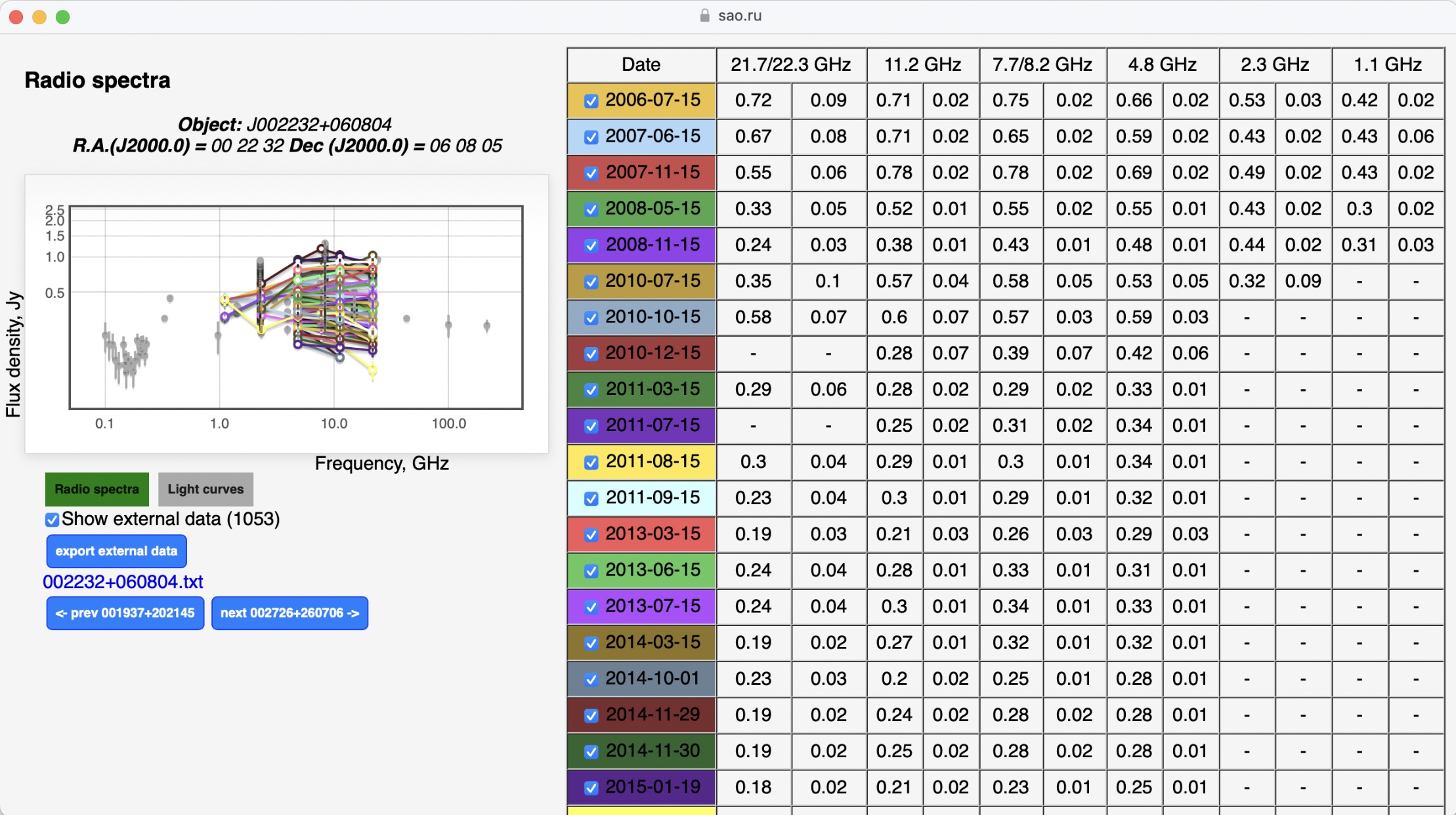}}
\caption{Multi-frequency radio spectra of the BL Lac object J0022$+$06, derived from the catalogue. There are 72 RATAN-600 observing epochs for the object from 2006.07 to 2021.01 (colored). The external radio data are shown by the grey circles.}
\label{fig:J0022_1}
\end{figure*}

\begin{figure*}
\centerline{\includegraphics[width=\textwidth]{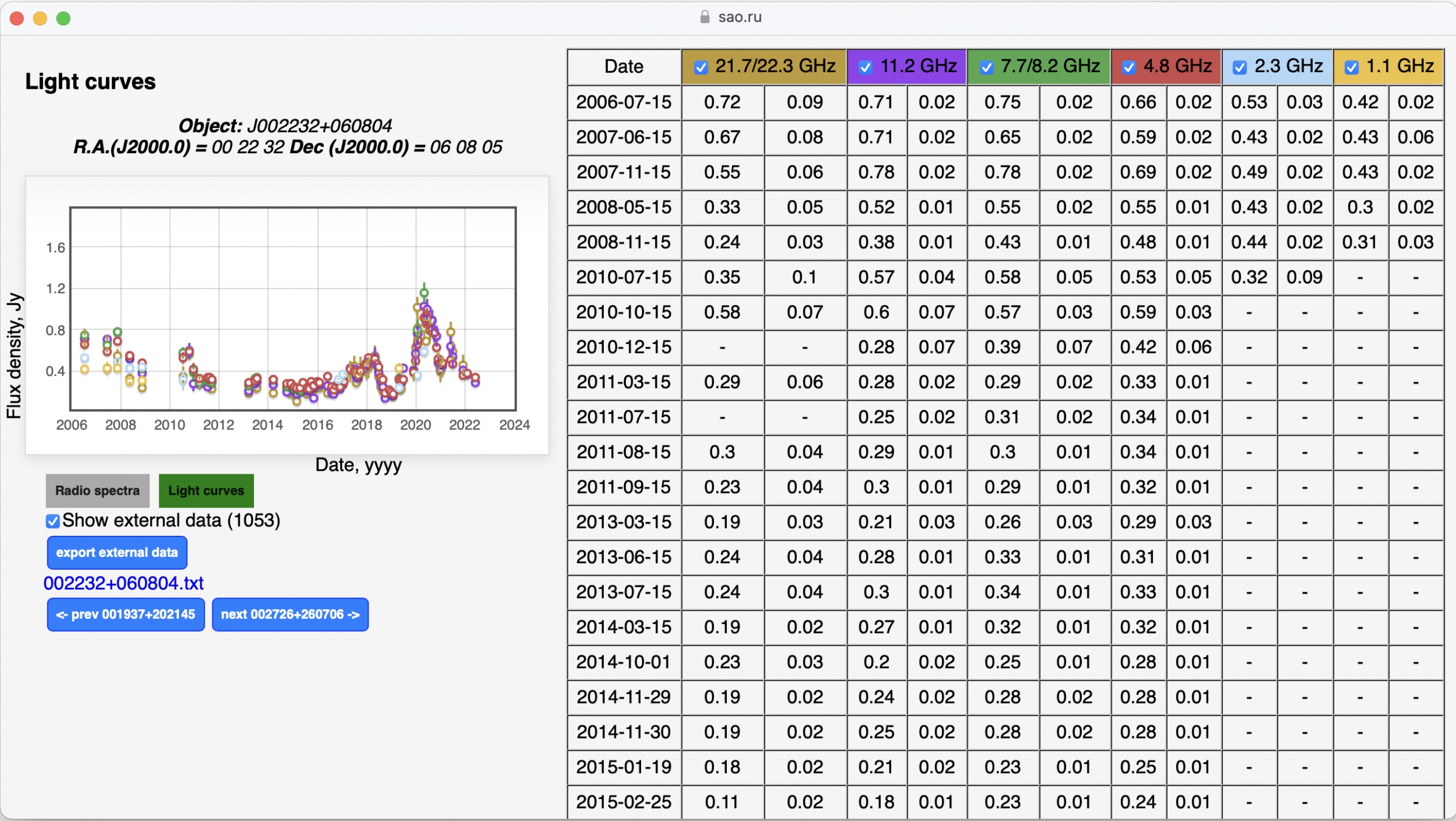}}
\caption{Multi-frequency light curves of the BL Lac object J0022$+$06, derived from the catalogue.}
\label{fig:J0022_2}
\end{figure*}

The catalogue is published as an interactive online tool to view and access the data at \url{https://www.sao.ru/blcat/}. The main page contains a list of objects with their characteristics placed in columns: the number of \mbox{RATAN-600} observing epochs for each object, name of the object (in the BZCAT5 format where it is possible), the J\,2000 right ascension (RA) and declination (Dec) coordinates, the redshift, the $R$-band magnitude, the averaged flux density at 4.7~GHz, the averaged luminosity at 4.7~GHz, and the type of the blazar (based on the Roma-BZCAT 5th edition). An example of the main table contents is given in Table~\ref{table3}. Each column can be sorted, and the columns containing the number of observing epochs, the redshift, the $R$-band magnitude, the flux density, the radio luminosity, and the blazar type also have distribution statistics. The radio data of any object/s can be accessed by clicking on the ``Data explorer'' option. It opens a new window with the \mbox{RATAN-600 data}, represented by the flux densities obtained at three to six frequencies quasi-simultaneously, the radio spectrum, and the light curves (an example is shown in Figs.~\ref{fig:J0022_1}~and~\ref{fig:J0022_2}). The radio spectrum could be supplemented with external data (``Show external data''), which can be exported right there (``Export external data'') in the format of a text file containing the blazar name, the frequency, the flux density, and the name of a catalogue or a bibliographic code to identify the literature.

\begin{table*}
\caption{\label{tab:param} 
The median and average values of the redshift $z$, flux density $S_{4.7}$, and radio luminosity $P_{4.7}$ at 4.7~GHz; here $N$ is the number of objects, the standard deviation for the average values is given in parentheses}
\centering
\begin{tabular}{l|c|c|c|c|c|c|c|c|c}
\hline
\multirow{2}{*}{Blazar type} &  \multicolumn{3}{|c|}{$z$}  &  \multicolumn{3}{|c|}{$S_{4.7}$, Jy}  & \multicolumn{3}{|c}{$P_{4.7}$, W Hz$^{-1}$}   \\
                             &   $N$  &  average & median & $N$ & average & median & $N$ & average & median\\
\hline
FSRQ       & 1097 & 1.47 (0.79) & 1.34 & 1097 & 0.59 (1.29) & 0.32 & 1065 & 5.0 (11.2)$\times10^{29}$ & 1.9$\times10^{29}$  \\ 
BL Lac     & 324  & 0.38 (0.35) & 0.27 & 539  & 0.27 (0.42) & 0.11 & 252  & 3.2 (9.7)$\times10^{28}$ & 1.9$\times10^{27}$ \\
Blaz.un.t. & 113  & 0.77 (0.71) & 0.58 & 128  & 0.79 (2.33) & 0.31 & 105  & 2.0 (2.5)$\times10^{29}$ & 3.5$\times10^{28}$ \\ 
\hline
\end{tabular}
\end{table*}

We provide spectral and variability indices, automatically calculated in different frequency intervals. The spectral index $\alpha$ defined from the power law $S_{\nu} \sim \nu^{\alpha}$, where $S_{\nu}$ is the flux density at frequency $\nu$, and $\alpha$ is a spectral slope. It is calculated using the standard formula: 
\begin{equation}
\label{sp:index}
\alpha=\dfrac{\log S_{2} - \log S_{1}}{\log{\nu}_{2} - \log{\nu}_{1}},         
\end{equation}
where $S_{1}$ and $S_{2}$ are the flux densities at frequencies $\nu_{1}$ and $\nu_{2}$, respectively. 

The variability index is estimated using the formula adopted from \citet{1992ApJ...399...16A}:
\begin{equation}
\label{variability}
{V}_{S}=\dfrac{(S_{i}-\sigma_{i})_{\rm max}-(S_{i}+\sigma_{i})_{\rm min}}
{(S_{i}-\sigma_{i})_{\rm max}+(S_{i}+\sigma_{i})_{\rm min}},
\end{equation}
where $S_{\rm max}$ and $S_{\rm min}$ are the maximum and minimum values of the flux density over all epochs of observations, while $\sigma_{\rm max}$ and $\sigma_{\rm min}$ are their root-mean-square errors. The spectral index can be calculated both for a single RATAN-600 observing epoch and for the entire period of observation as an averaged index. Also we provide the estimates of the monochromatic radio luminosity calculated at 4.7~GHz following the equation:
\begin{equation}
P_{4.7} = 4 \pi D_{L}^2 S_{4.7} (1+z)^{-\alpha -1},
\end{equation}
where $S_{4.7}$ is the measured flux density at 4.7~GHz, $z$ is the redshift, $\alpha$ is the average spectral index at 4.7~GHz, and $D_{L}$ is the luminosity distance. For calculation of luminosity distances, we used the $\Lambda$CDM cosmology with $H_0 = 67.74$~km\,s$^{-1}$\,Mpc$^{-1}$, $\Omega_m=0.3089$, and $\Omega_\Lambda=0.6911$ \citep{2016A&A...594A..13P}. The uncertainties of the parameters ($\alpha$, ${V}_{S}$, and $P_{4.7}$) are calculated as the square root of the sum of the squared errors of the parameters included in the formula.

\begin{table*}
\caption{\label{tab:ind} 
Theaverage values of spectral indices at 11.2--22.3, 8.2--11.2,  4.7--8.2, 2.3--4.7, and 1.2--2.3~GHz for three blazar types. The standard deviation for average spectral indices is indicated in parentheses}
\centering
\begin{tabular}{l|c|c|c|c|c}
\hline
Blazar type & $\alpha_{11.2-22.3}$ & $\alpha_{8.2-11.2}$ & $\alpha_{4.7-8.2}$ & $\alpha_{2.3-4.7}$ & $\alpha_{1.2-2.3}$ \\
\hline
FSRQ       & $-$0.09 (0.41) & $-$0.22 (0.44) & $-$0.13 (0.45) & $+$0.16 (0.38) & $-$0.11 (0.52)  \\ 
BL Lac     & $+$0.03 (0.46) & $-$0.21 (0.54) & $-$0.17 (0.52) & $-$0.17 (0.60) & $-$0.46 (0.64)  \\
Blaz.un.t. & $-$0.10 (0.48) & $-$0.31 (0.49) & $-$0.21 (0.56) & $-$0.01 (0.55) & $-$0.22 (0.70) \\
\hline
\end{tabular}
\end{table*}

The content of the main table, measured RATAN-600 flux densities, and calculated parameters ($\alpha$, ${V}_{S}$, and $P_{4.7}$) could be downloaded (e.g., as a \textit{CSV} format text file) using the ``Export main table and RATAN-600 data'' button. On the website there is also the ``Help'' button, where the description of the catalogue, data, and all the tools are given.

\section{THE RADIO PROPERTIES OF BLAZARS IN THE CATALOGUE}
\label{properties}

This section briefly describes some radio properties of the blazar subsamples in the catalogue. The spectral indices and radio variability were calculated based on observed data obtained using one instrument and at four--six frequencies simultaneously, which minimizes the influence of measurement inhomogeneities on the result. The median and average values of the redshift, flux density, and radio luminosity at 4.7~GHz for FSRQs, BL Lacs, and Blazars of uncertain type are presented in Table~\ref{tab:param}. The obtained radio properties of the subsamples reflect a wide range of blazar parameters and statistical differences between them.

\subsection{Spectral Index}
We analysed the averaged radio spectra of blazars compiled from the RATAN-600 quasi-simultaneous measurements over a time period from 2005 to 2022. The spectral indices $\alpha$ were calculated using Equation \ref{sp:index}. Their median and average values at five frequency intervals are listed in Table~\ref{tab:ind} for three blazar subsamples. The median values of $\alpha_{8.2-22.3}$ are $-0.15$, $-0.08$, and $-0.18$ for the FSRQ, BL Lac, and Blaz.un.t. types respectively. For $\alpha_{2.3-4.7}$ the median values are equal to $+0.14$, $-0.08$, and $+0.01$ respectively. 

\begin{figure}
\centerline{\includegraphics[width=\columnwidth]{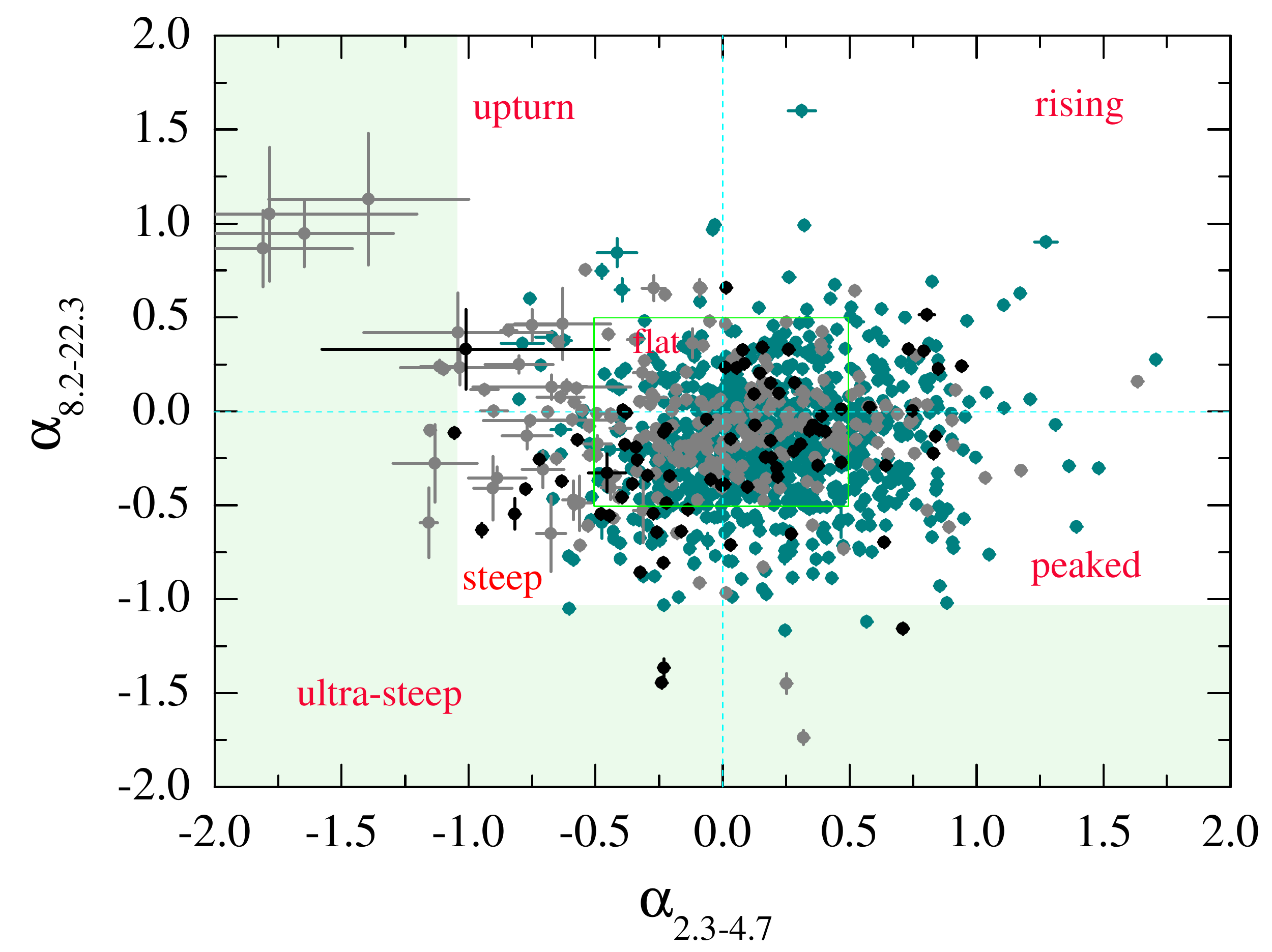}}
\caption{Two-colour diagram. The green square indicates the spectral index range from $-0.5$ to $0.5$, the blue lines mark the zero spectral indices, and the ultra-steep spectrum sources area is presented with a light green colour. FSRQs are marked by green colour, BL Lacs are grey, and blazars of an uncertain type are marked by black colour.}
\label{fig7}
\end{figure}

Figure~\ref{fig7} presents the spectral type distribution in a convenient form of a two-colour diagram in which the low-frequency spectral index is plotted against the high-frequency index. The green square corresponds to the spectral index range from $-0.5$ to $0.5$ (``blazar box''). The light green color indicates the area of ultra-steep spectra with $\alpha \leq -1.1$. The other areas are marked according to the spectral classification criteria from Table~\ref{tab:type}.

The vast majority of blazars have a flat (876, 50.5\%) or rising (476, 27.5\%) radio spectrum. We found a small part of blazars (26, 1.5\%) having peaked radio spectra with a turnover in their average radio spectrum (Table~\ref{tab:type}) and spectral indices $\alpha_{\rm low}\ge+0.5$ and $\alpha_{\rm high} \leq-0.5$, which correspond to the slopes of classical gigahertz-peaked spectrum (GPS) sources \citep{1997A&A...321..105D}. A very small number of these blazars (2.2\%) have ultra-steep radio spectra with $\alpha \leq -1.1$. 

\begin{table}
\caption{\label{tab:type}Spectral types in the sample. N - is the number of sources.}
\centering
\begin{tabular}{l|c|@{~}r@{~}l}
\hline
Type & Criteria  &   \multicolumn{2}{c}{$N$ (\%)} \\
\hline
Flat    & $-0.5 \leq \alpha \leq 0$  &  876 &(50.5) \\ 
Peaked  & $\alpha_{\rm low}\ge+0.5$, $\alpha_{\rm high} \leq-0.5$ & 26 &(1.5)  \\
Rising  & $\alpha>0$ &  476 &(27.5) \\ 
Upturn  & $\alpha_{\rm low}<0$, $\alpha_{\rm high}>0$ & 109 &(6.3) \\ 
Steep  & $-1.1 < \alpha < -0.5$ & 208& (12) \\ 
Ultra-steep  & $\alpha \leq -1.1$ & 39& (2.2) \\ 
\hline
\end{tabular}
\end{table}

\subsection{Variability}
\begin{figure*}
\centerline{\includegraphics[width=\textwidth]{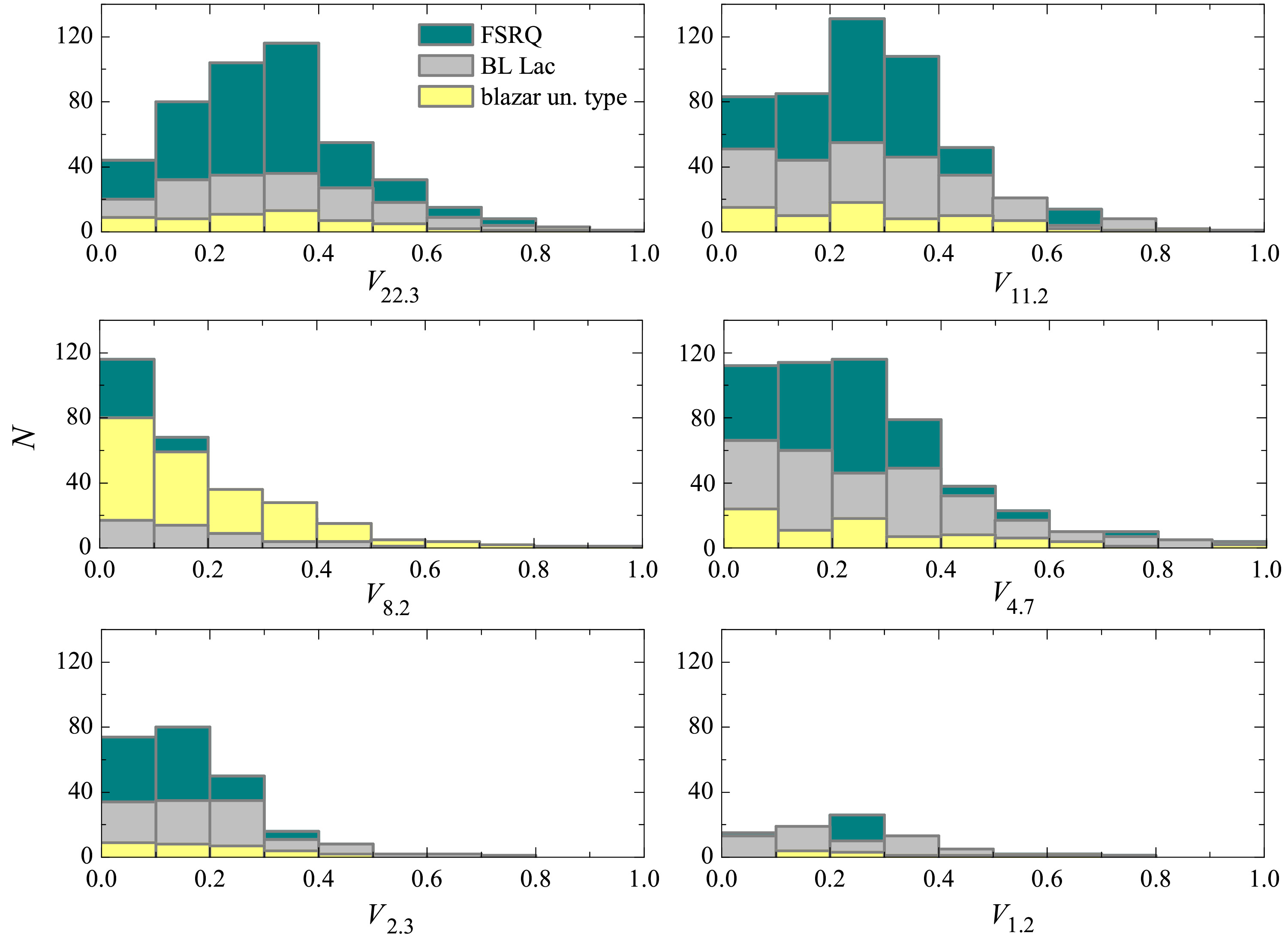}}
\caption{The variability index distributions at 22.3, 11.2, 8.2, 4.7, 2.3, and 1.2~GHz for the subclasses of blazars from the catalogue.}
\label{var}
\end{figure*}

The variability indices $V_{S}$ (Equation~\ref{variability}) were calculated using quasi-simultaneous RATAN-600 measurements at frequencies of 2.3, 4.7, 7.7/8.2, 11.2, and 21.7/22.3~GHz. The $V_{S}$ distributions for different subclasses are shown in Fig.~\ref{var}. The average values of $V_{S}$ for each type are shown in Table~\ref{tab:var1}. The higher variability indices was found at 22.3~GHz. On average, the BL Lac
blazars are more variable at all frequencies. The average and median $V_{S}$ values at 22.3~GHz span from 0.07 to 0.30 with maximum values of 1.77, 0.90, and 0.89 for FSRQ, BL Lac, and Blaz.un.t.respectively. The maximum variability indices 0.90--1.77 were estimated for FSRQ and BL Lac blazars at 4.7--22.3~GHz. For the Blaz.un.t. subsample the maximum variability indices are slightly lower at almost all frequencies, $V_{\rm max}$=0.53--0.89. 

\begin{table*}
\caption{\label{tab:var1} 
The average values of the variability indices $V_{S}$ calculated at five frequencies, 2.3, 4.7, 8.2, 11.2 and 22.3~GHz, for three blazar types. The standard deviation for average variability indices is indicated in parentheses}
\centering
\begin{tabular}{l|c|c|c|c|c}
\hline
Blazar type  & $V_{22.3}$ & $V_{11.2}$ & $V_{8.2}$ & $V_{4.7}$ & $V_{2.3}$ \\
\hline
FSRQ       & 0.28 (0.21) & 0.23 (0.23) & 0.10 (0.14) & 0.23 (0.20) & 0.15 (0.12) \\ 
BL Lac     & 0.30 (0.23) & 0.27 (0.20) & 0.17 (0.18) & 0.25 (0.25) & 0.17 (0.15) \\
Blaz.un.t. & 0.26 (0.22) & 0.26 (0.20) & 0.15 (0.14) & 0.15 (0.14) & 0.16 (0.14) \\
\hline
\end{tabular}
\end{table*}

\section{Summary}
\label{summary}
We present the online ``RATAN-600 multi-frequency catalogue of blazars'' (BLcat) that summarises radio continuum measurements of more than 1700 objects in the frequency range from tens of MHz up to hundreds of GHz. The quasi-simultaneous data at 1.2--22.3~GHz are represented by the RATAN-600 measurements since 2005, and  additional literature data from external sources cover a time period of 30--40~years. 

The catalogue provides estimation of flux densities, spectral indices, variability, and radio luminosity for the large blazars list. It is a useful tool for analysing the flare activity and synchrotron radio spectra of AGNs. The catalogue is regularly updated with new RATAN-600 and literature measurements. 

The users who will make use of the \mbox{RATAN-600} multi-frequency catalogue of blazars in a publication are kindly requested to acknowledge the source of the data by referencing this paper.

\section*{Acknowledgements}
This work is supported in the framework of the national project ``Science'' by the Ministry of Science and Higher Education of the Russian Federation under the contract 075-15-2020-778. The observations were carried out with the RATAN-600 scientific facility. Observations with RATAN-600 are supported by the Ministry of Science and Higher Education of the Russian Federation. This research has made use of the CATS database, operated at SAO RAS, Russia. This research has made use of the Roma-BZCAT NASA/IPAC Extragalactic Database (NED), which is operated by the Jet Propulsion Laboratory, California Institute of Technology, under contract with the National Aeronautics and Space Administration. This research has made use of the SIMBAD database, operated at CDS, Strasbourg, France


\bibliographystyle{mnras} %
\bibliography{blcat} %

\bsp	
\label{lastpage}
\end{document}